\documentclass[amsmath,twocolumn,superscriptaddress,prl,aps]{revtex4}

\usepackage{amsmath}

\usepackage{epsfig}
\usepackage{graphicx}
\usepackage{dcolumn}
\usepackage{bm}
\usepackage{amssymb}

\begin{document}


\title{$\mathcal{PT}$-symmetric phase in kagome photonic lattices}

\author{Gia-Wei Chern}
\affiliation{Department of Physics, University of Virginia, Charlottesville, VA 22904}
\affiliation{Theoretical Division and Center for Nonlinear Studies, Los Alamos National Laboratory, Los Alamos, NM 87545}

\author{Avadh Saxena}
\affiliation{Theoretical Division and Center for Nonlinear Studies, Los Alamos National Laboratory, Los Alamos, NM 87545}

\date{\today}

\begin{abstract}
Kagome lattice is a two-dimensional network of corner-sharing triangles and is often associated with geometrical frustration. In particular, the frustrated coupling between waveguide modes in a kagome array leads to a dispersionless flat band consisting of spatially localized modes. Here~we propose a complex photonic lattice by placing $\mathcal{PT}$-symmetric dimers at the kagome lattice points. Each dimer corresponds to a pair of strongly coupled waveguides. With balanced arrangement of gain and loss on individual dimers, the system exhibits a $\mathcal{PT}$-symmetric phase for finite gain/loss parameter up to a critical value. The beam evolution in this complex kagome waveguide array exhibits a novel oscillatory rotation of optical power along the propagation distance. Long-lived local chiral structures originating from the nearly flat bands of the kagome structure are observed when the lattice is subject to a narrow beam excitation.
\end{abstract}


\maketitle

Photonic lattices composed of balanced gain and loss waveguides~\cite{ganainy07,guo09,ruter10} have attracted considerable attention due of their potential applications in optical beam engineering and image processing~\cite{makris08,ramezani12a,bender15,gu15}. These photonic lattices belong to a larger class of intriguing metamaterials that exhibit the parity-time ($\mathcal{PT}$) symmetry~\cite{regensburger12,feng13,peng14}. The notion of $\mathcal{PT}$-symmetry was originally introduced as a complex extension of quantum mechanics~\cite{bender98}. It was shown that non-Hermitian Hamiltonians commuting with the combined $\mathcal{PT}$ operator may exhibit entirely real energy spectra. In optics, this combined symmetry requires that the complex refractive index obeys the condition $n(\mathbf r) = n^*(-\mathbf r)$. The system is then in the so-called exact phase with real propagation constants for a small gain and loss coefficient. As this coefficient increases, the system undergoes a transition into a  phase with broken $\mathcal{PT}$-symmetry. Although optical realization of $\mathcal{PT}$-symmetric Hamiltonians is usually based on the formal analogies between Schr\"odingier and paraxial Helmholtz equations, it was demonstrated recently that  $\mathcal{PT}$-symmetric optical phases exist in regimes well beyond the paraxial approximation~\cite{huang14}. 

Motivated by recent experimental developments~\cite{guo09,ruter10}, $\mathcal{PT}$-symmetric photonic lattices have been theoretically studied in various geometries, including the one-dimensional (1D) chain~\cite{makris08,longhi09} and ladder~\cite{bendix10}, as well as 2D square~\cite{makris08}, honeycomb~\cite{szameit11,ramezani12b}, and triangular lattices~\cite{wang15}. Several intriguing features have been observed including, among others, power oscillation, double diffraction, and a new type of conical diffraction. Inclusion of nonlinearity also leads to several new soliton solutions~\cite{zeng12,wang15,musslimani08,dmitriev10}. Another lattice structure of both fundamental interest and practical application is the kagome lattice. The interest in it is partly due to the optical flat bands that have been recently demonstrated in the so-called Lieb lattice~\cite{vicencio15,mukherjee15}. In both kagome and Lieb lattices, strong geometrical frustration leads to a degenerate flat band in their coupled-mode spectra. In particular, kagome is one of the most extensively studied structure in highly frustrated magnets~\cite{kagome}. The eigenmodes of the flat band are highly localized spatially even in the linear regime. Diffraction-free transmission can be achieved in such lattices by decomposing a given image into the local modes~\cite{vicencio14}. The significantly reduced dispersion of the flat band also indicates enhanced nonlinear effects~\cite{vicencio13,law09}. 
In addition to the flat band, the kagome spectrum also contains several topologically nontrivial band-crossing points; gapping out these points could lead to novel optical topological phases~\cite{sun09}.

Our goal in this paper is to present a $\mathcal{PT}$-symmetric kagome lattice which can be used to explore the intriguing interplay of flat band and novel phenomena due to complex potentials. However, a straightforward arrangement of gain and loss in a kagome lattice fails to produce a $\mathcal{PT}$-symmetric phase. Our strategy is to build the complex waveguide array based on $\mathcal{PT}$-symmetric dimers, which have been shown to play an important role in realizing an exact $\mathcal{PT}$-symmetric phase in other structures~\cite{dmitriev10,bendix10}. A $\mathcal{PT}$-symmetric dimer is a pair of strongly coupled waveguides with balanced gain and loss $\pm i \gamma$. In fact, optical lattices built from such dimers possess a local $\mathcal{PT}$ symmetry associated with each dimer and are rather robust against disorder~\cite{bendix10}. On the other hand, the dimerization also significantly lowers the lattice symmetry to no more than a mirror reflection. Indeed, in order to realize the exact $\mathcal{PT}$ phase, the $C_3$ symmetry is lost in the complex triangular and honeycomb lattices proposed in Refs.~\cite{szameit11,ramezani12b,wang15}.

In this paper, we propose and study a kagome photonic lattice built from $\mathcal{PT}$-symmetric dimers. The lattice structure can be viewed as two interpenetrating twisted kagome lattices and retains a full $C_3$ point group symmetry. We show that an exact $\mathcal{PT}$-symmetric phase exists for gain/low parameter smaller than a critical value $\gamma_c$. More importantly, the band structure with $\gamma < \gamma_c$ contains two nearly flat bands inherited from the kagome tight-binding spectrum. The $\mathcal{PT}$-symmetry breaking transition occurs when the two nearly flat bands collapse at the Brillouin zone center. Our linear beam dynamics simulations uncover interesting oscillatory rotations of optical power along the propagation distance. In particular, we find long-lived local chiral structures consisting of quasi-local modes when the lattice is subject to a narrow beam excitation. The non-reciprocal nature of the chiral beam transmission is also discussed. 

\begin{figure}
\includegraphics[width=0.8\columnwidth]{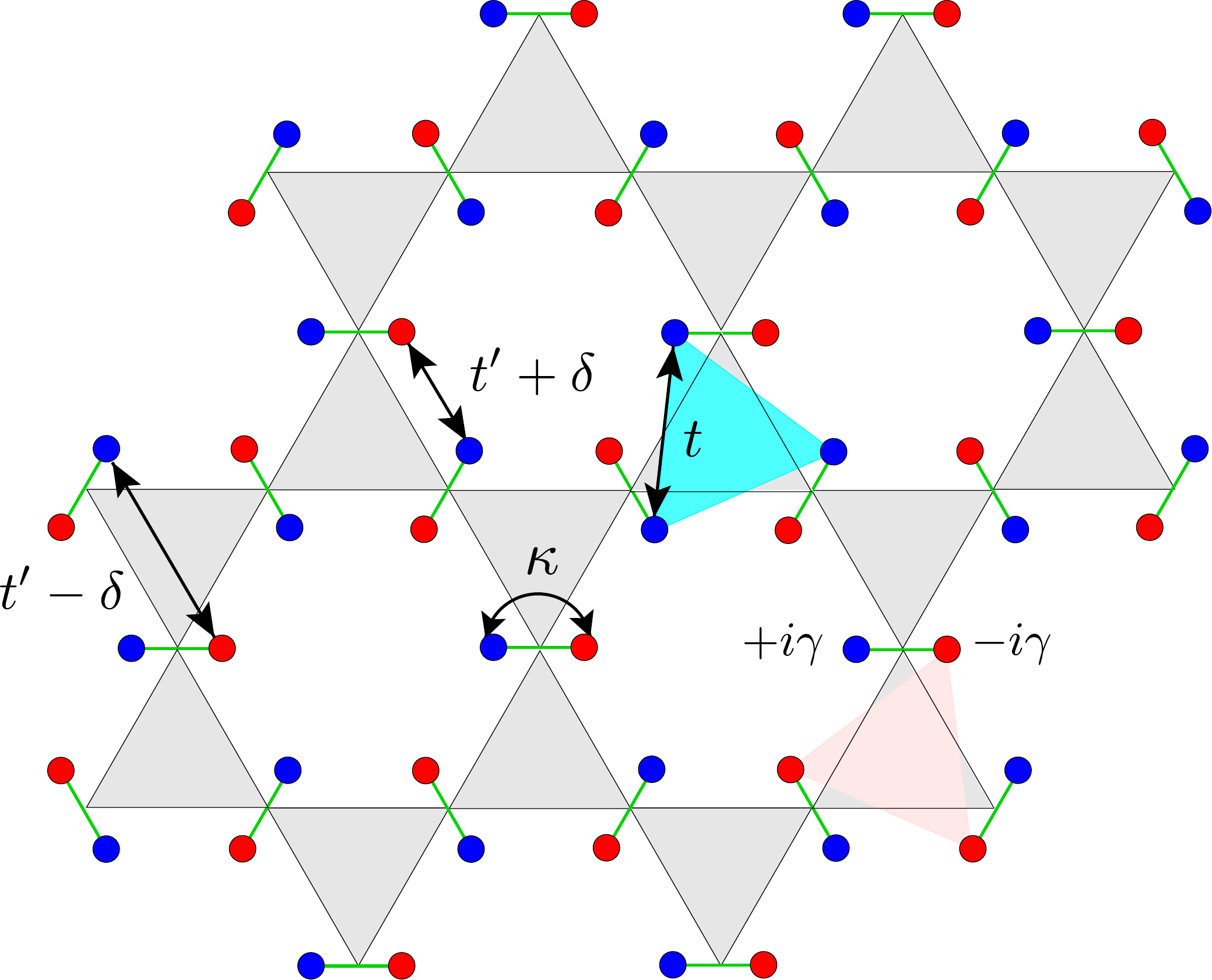}
\caption{\footnotesize (Color online) kagome photonic lattice consisting of $\mathcal{PT}$-symmetric dimers. The blue and red circles denote waveguides with a gain and a loss parameter $\pm i \gamma$, respectively. 
$\kappa$ is the intra-dimer coupling, while $t$ and $t' \pm \delta$ denote the nearest-neighbor (NN) couplings between same and different sublattices, respectively. 
\label{fig:kagome}}
\end{figure}

The complex kagome lattice depicted in Fig.~\ref{fig:kagome} consists of dimers placed at vertices of each triangle. Since the kagome-lattice sites correspond to bonds in the dual honeycomb lattice, the same structure can be obtained by placing dimers on the bonds of the honeycomb lattice, with dimer orientations perpendicular to the bonds. Each dimer consists of two types of waveguides: type $A$ is made of gain material, whereas type $B$ exhibits an equal amount of loss. Interestingly, the $A$ sublattice itself forms a twisted kagome lattice, while the $B$ waveguides are arranged in another twisted kagome with opposite chirality.  We assume that each waveguide supports only one mode, while light is transferred between neighboring waveguides through optical tunneling. In the tight-binding description~\cite{joglekar13}, the diffraction dynamics of the optical field amplitude $\Psi_n = (a_n, b_n)^T$ at the $n$-th dimer evolves according to the following coupled-mode equation:
\begin{eqnarray}
	\label{eq:coupled-mode}
	i \frac{d a_n}{dz} = +i \gamma\, a_n + \kappa\, b_n + \sum_m^{\rm NN} \left[ t\, a_m +  (t' \pm \delta) \, b_m \right], \nonumber \\
	i \frac{d b_n}{dz} = -i \gamma\, b_n + \kappa\, a_n + \sum_m^{\rm NN} \left[ t\, b_m + (t' \pm \delta) \, a_m \right].
\end{eqnarray}
Here $\gamma$ is the gain/loss parameter, $\kappa$ is the dominant intra-dimer coupling, $t$ is the nearest-neighbor (NN) coupling constant between waveguides of same sublattices, $t' \pm \delta$ denote the two distinct NN couplings between different sublattices; see Fig.~\ref{fig:kagome}. The summation above is restricted to the NN pairs. 

\begin{figure}
\includegraphics[width=0.99\columnwidth]{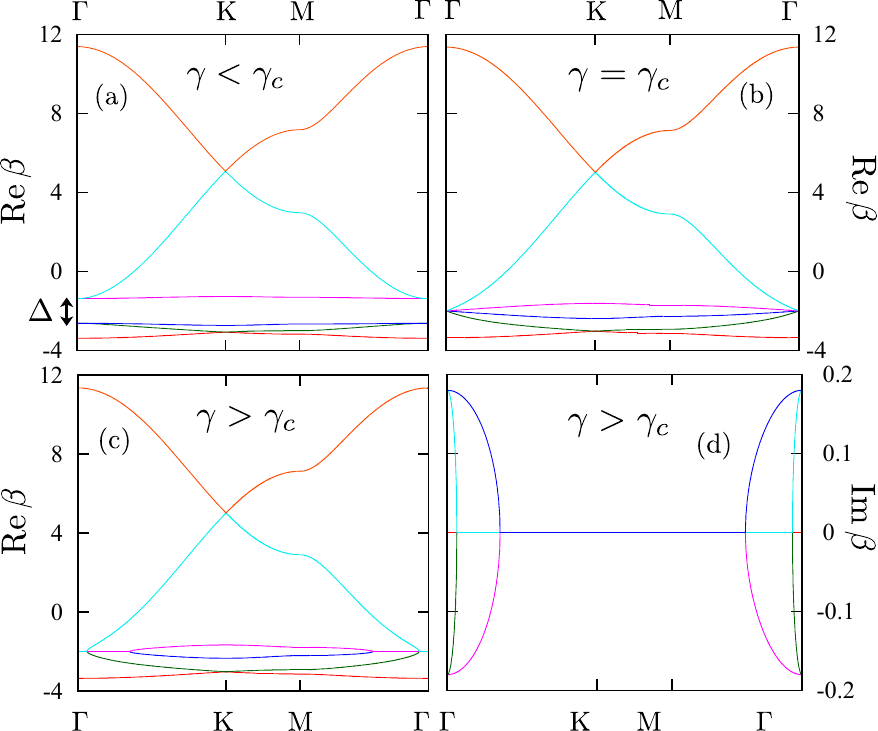}
\caption{\footnotesize (Color online) Band structure of the kagome dimer lattice along high-symmetry directions for varying gain and loss parameter $\gamma$. The coupling constants (in units of the NN coupling $t$) are $\kappa = 3$, $t' = 1.1$, and $\delta = 0.1$. The critical $\gamma_c = 0.8$. (a) and (b) show the (completely) real spectra when $\gamma = 0.7$ and $\gamma = \gamma_c$, respectively. (c) and (d) show the real and imaginary parts, respectively, of the eigen-spectrum for $\gamma = 1.2$.
\label{fig:spectrum}}
\end{figure}

We next examine the dispersion relations of the complex kagome lattice. To this end, we  note that the kagome site index can be represented as $n = (s, \mathbf r_n)$, where $s = 1, 2, 3$ indicates the three sublattices of kagome, and $\mathbf r_n$ denotes the Bravais lattice point of the underlying triangular lattice. After introducing a Fourier transform for field amplitude $a_n = (1/\sqrt{N}) \sum_{\mathbf k} a_{s, \mathbf k} \exp(i \mathbf k \cdot \mathbf r_n)$ and a similar expression for $b_n$, the coupled-mode equations become $i d\Psi_{\mathbf k}/dz = \mathcal{H}_{\mathbf k} \,\Psi_{\mathbf k}$, where
\begin{eqnarray}
	\mathcal{H}_{\mathbf k} = \left( \begin{array}{cc} \mathcal{A}_{\mathbf k} & \mathcal{B}_{\mathbf k} \\
	\mathcal{B}_{\mathbf k}^\dagger & \mathcal{A}_{\mathbf k}^\dagger \end{array} \right),
\end{eqnarray}
and the two $3\times 3$ matrices $\mathcal{A}$ and $\mathcal{B}$ are
\begin{eqnarray}
	\mathcal{A}_{\mathbf k} = \left( \begin{array}{ccc}
	i \gamma &  t  c_3 &   t  c_2 \\
	 t  c_3 & i \gamma &  t  c_1 \\
	 t  c_2 &  t  c_1 & i \gamma 
	\end{array} \right), 
\end{eqnarray}
\begin{eqnarray}
	\mathcal{B}_{\mathbf k} = \left( \begin{array}{ccc}
	\kappa &  t' c_3 -  i \delta s_3 &  t' c_2 -  i \delta s_2 \\
	 t' c_3 + i \delta s_3 & \kappa & t' c_1 + i \delta s_1 \\
	 t' c_2 + i \delta s_2 &  t' c_1 -  i \delta s_1 & \kappa 
	\end{array} \right).
\end{eqnarray}
For convenience,  we have introduced factors $c_i = 2 \cos(\mathbf k \cdot \mathbf d_i)$ and $s_i = 2 \sin(\mathbf k \cdot \mathbf d_i)$  ($i = 1, 2, 3$), the three vectors $\mathbf d_{1, 2} = a (1, \mp \sqrt{3})/4$, $\mathbf d_3 = a (1/2, 0)$ connect the neighboring sites on kagome, and $a$ is the lattice constant. The dispersion relations of the  eigenmodes are given by the stationary solutions $\Psi_{\mathbf k}  \sim \exp(i \beta z)$, where the propagation constant $\beta = \beta(\mathbf k)$ is the eigenvalue of~$\mathcal{H}_{\mathbf k}$.

\begin{figure}
\includegraphics[width=0.99\columnwidth]{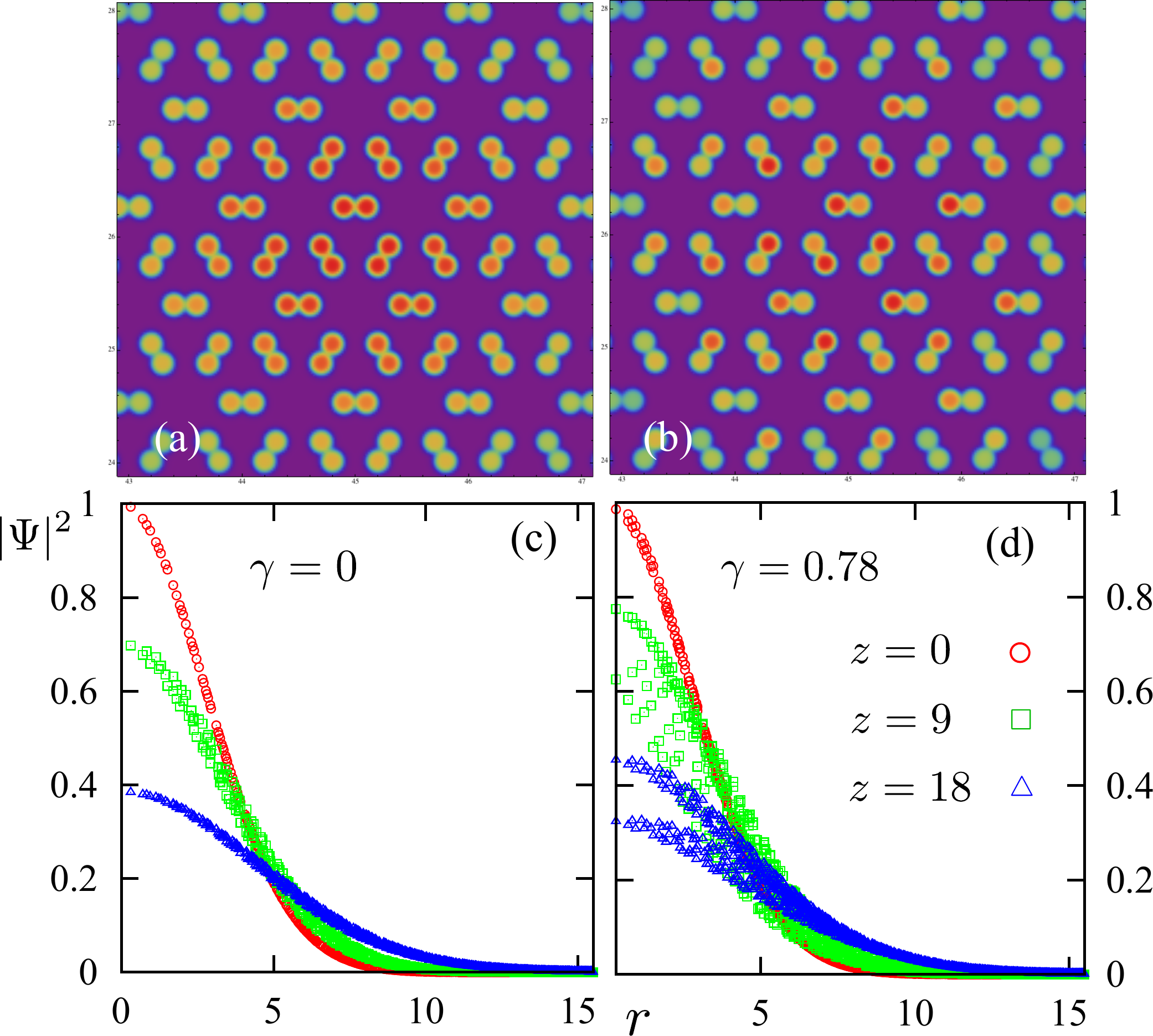}
\caption{\footnotesize (Color online) Snapshots of the intensity profile~$|\Psi_n|^2$ under wide beam excitations for (a) a passive array ($\gamma = 0$) and (b) a complex array with $\gamma = 0.75$; the width of the incident beam is $w_0 = 5.5 a$. The lower panels show the radial distribution of optical power for (c)~a passive and (d) a complex $\gamma = 0.75$ kagome array.
\label{fig:sim1}}
\end{figure}

As is well established in other non-Hermitian systems, the $\mathcal{PT}$ symmetry is a necessary but not a sufficient condition for the reality of the eigenvalues $\beta$. By using spectral techniques we numerically determined the existence of a $\mathcal{PT}$ threshold $\gamma_c$, below which the propagation constants of all bands and wave vectors are real. The critical $\gamma_c$ depends on coupling constants $\kappa$ and $t'$, to be discussed below. Above this threshold, the system undergoes a transition into a phase with partially complex eigenvalues $\beta$. Fig.~\ref{fig:spectrum}(a) shows the band structure of the kagome lattice when $\gamma < \gamma_c$. This spectrum inherits several features characteristic of the kagome tight-binding model; it can be viewed as two (scaled) copies of the kagome spectrum. First, there are two pairs of Dirac points located at the corners of the Brillouin zone~\cite{vicencio13}. Second, there are two quadratic band crossing points at the zone center. These quadratic crossing points are topologically nontrivial in the sense that each of them carries a $2\pi$ Berry flux~\cite{sun09}. Finally, there are two nearly flat bands (the 3rd and 4th) due to geometrical frustration; the weak dispersion is caused by the coupling between the two twisted kagome sublattices $A$ and $B$. 

The two nearly flat bands are separated by a finite gap $\Delta$ as shown in Fig.~\ref{fig:spectrum}(a). With increasing gain/loss parameter, the closing of the gap at $\mathbf k = 0$ coincides with the $\mathcal{PT}$-symmetry breaking transition. Analytical calculation finds a gap $\Delta_{\Gamma} = 2 \sqrt{(\kappa - 2t')^2 - \gamma^2}$ at the $\Gamma$ point. Setting $\Delta_{\Gamma} =0$ gives the critical value for the $\mathcal{PT}$ transition: $\gamma_c = \kappa - 2 t'$, which is independent of $t$ and $\delta$. The two topological quadratic band-crossing points merge at the Brillouin zone center when $\gamma = \gamma_c$; see Fig.~\ref{fig:spectrum}(b). The $\mathcal{PT}$ phase transition here exhibits all the characteristics of an exceptional-point singularity. Other than the coalescing of eigenmodes, signaling a collapsed Hilbert space, we have also confirmed numerically a divergent Petermann factor~\cite{zheng10,petermann79} when $\gamma \to \gamma_c$.

\begin{figure}
\includegraphics[width=0.99\columnwidth]{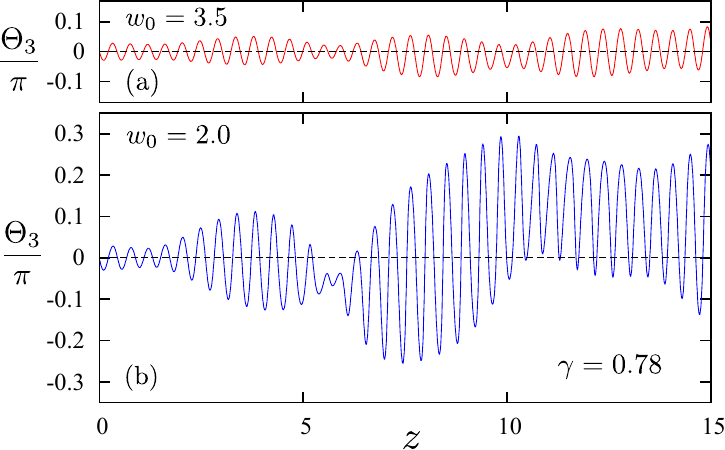}
\caption{\footnotesize (Color online) The oscillatory power rotations in a $\mathcal{PT}$-symmetric kagome lattice ($\gamma = 0.78$) for (a) wide and (b) narrow beam excitations. 
\label{fig:sim2}}
\end{figure}


The complex guiding potentials also have profound effects on linear beam dynamics of the kagome waveguide arrays. By exciting the $\mathcal{PT}$-symmetric lattice with a Gaussian beam at $z =0$, we numerically integrate Eq.~(\ref{eq:coupled-mode}) for a large system with up to $6\times 10^4$ waveguides. We first consider the situation when the array is excited by a wide beam of width $w_0 = 5.5 a$; the center of the beam coincides with one of the triangle center. For a passive kagome lattice, the beam evolution follows the standard diffraction pattern as shown in Fig.~\ref{fig:sim1}(a). The envelop of the optical field remains circularly symmetric with respect to the beam center, while its amplitude slowly decays with distance; see Fig.~\ref{fig:sim1}(c). On the other hand, the intensity profile, Fig.~\ref{fig:sim1}(b), of a complex array with $\gamma = 0.75$ exhibits a rather inhomogeneous power distribution within individual dimers.  As the optical power oscillates in the kagome lattice, the radial profile of $|\Psi|^2$ shows multiple branches as shown in Fig.~\ref{fig:sim1}(d). Detailed examination reveals that the optical field rotates periodically about the beam center. To quantify this oscillatory rotation, we first compute the angular Fourier components of the intensity profile: $\mathcal{F}_m \equiv \sum_n |\Psi_n|^2\, \exp(i m\, \theta_n)$, where $m$ is an integer, and $\theta_n = {\rm arctan}(y_n/x_n)$ is the angular coordinate of the $n$th waveguide. The fact that the optical field maintains a $C_3$ symmetry indicates that the first nontrivial higher harmonic is $m = 3$. We can then define an angle $\Theta_3 \equiv {\rm arctan}[{\operatorname{Im}}(F_3) /{\operatorname{Re}}(F_3)]$ to measure the shift of the optical power from the $C_3$ symmetric axis of the lattice. For a passive lattice, this angle  remains zero throughout  the beam propagation. On the contrary, $\Theta_3$ oscillates with $z$ in the exact $\mathcal{PT}$-symmetric phase of the kagome lattice; see Fig.~\ref{fig:sim2}(a). This result is reminiscent  of the power oscillation phenomena of a single $\mathcal{PT}$ dimer.

\begin{figure}[t]
\center
\includegraphics[width=0.85\columnwidth]{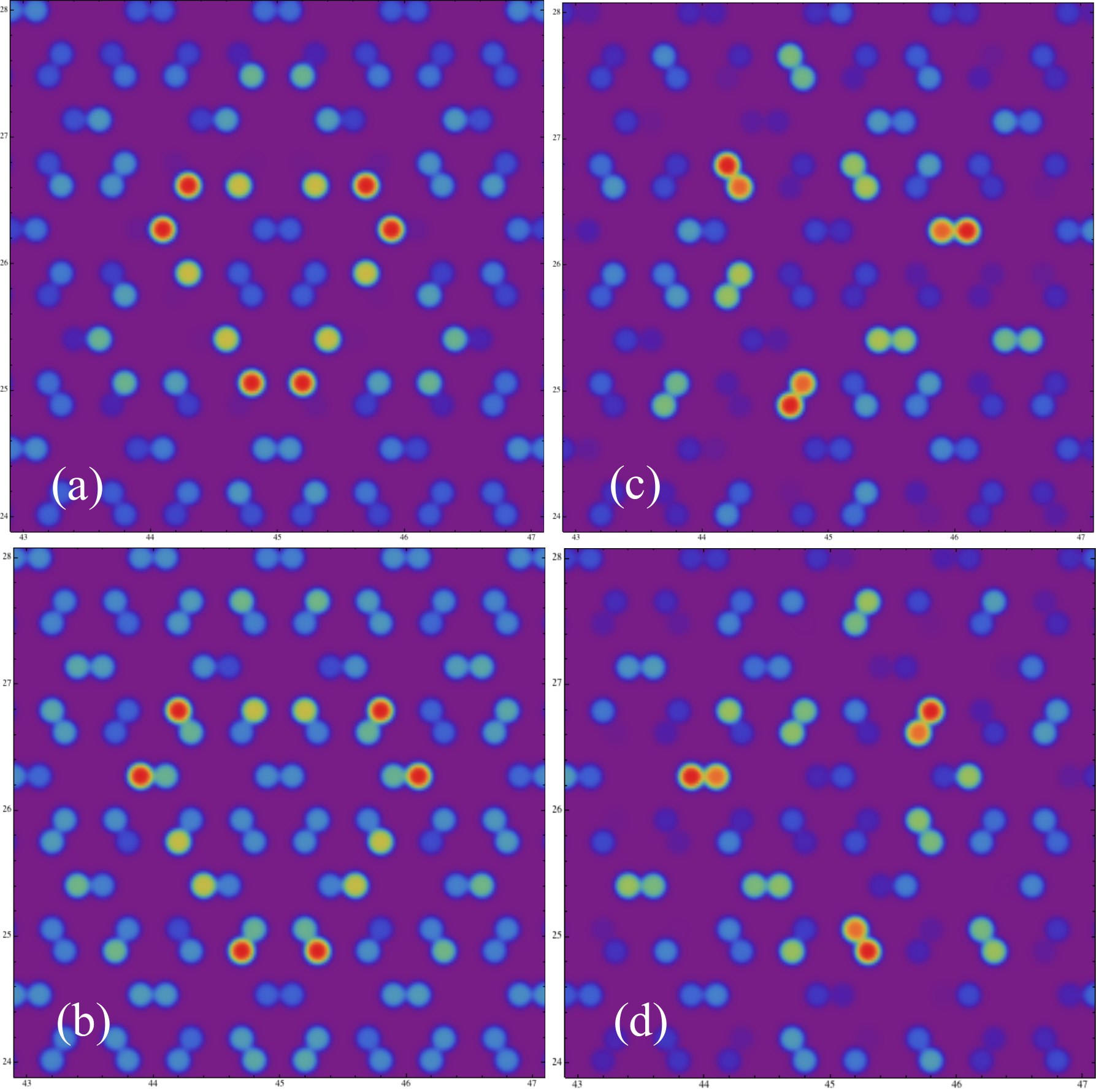}
\caption{\footnotesize (Color online) Snapshots of the quasi-stationary local structure under narrow beam excitations ($w_0 = a$) for (a), (b) a passive array ($\gamma = 0$) and (c), (d) a complex array with $\gamma = 0.75$.
\label{fig:sim3}}
\end{figure}

An input beam with a narrow width excites a broad range of eigenmodes including those in the nearly flat bands. After the optical power spreads over the lattice through the dispersive modes, a long-lived local structure remains in the central region of the input beam. These local structures are quasi-stationary. For example, in a passive array, the intensity profile oscillates between the two patterns shown in Fig.~\ref{fig:sim3}(a) and (b); the pattern repeats itself with a period $\Delta z \approx 3.5$. The introduction of gain/loss to the waveguides gives rise to a {\em chiral} local structure; two snapshots of such local patterns are shown in Fig.~\ref{fig:sim3}(c) and (d) for a $\mathcal{PT}$-symmetric kagome lattice with $\gamma = 0.78$. The dynamical evolution of the local structures is more complex than that of a passive array. Again, while maintaining a $C_3$ symmetry, the chiral pattern rotates about the beam axis. This is illustrated in Fig.~\ref{fig:sim2}(b) which shows the angle $\Theta_3$ as a function of propagation distance $z$ for a narrow beam excitation. 
The chiral nature of the local structure also manifests itself in the beam dynamics: the moving average of $\Theta_3$ deviates significantly from zero especially at large $z$.

To summarize, we have presented a $\mathcal{PT}$-symmetric kagome photonic lattice. The complex waveguide array supports an exact $\mathcal{PT}$-symmetric phase for gain/loss parameter below a finite threshold. The eigen-spectrum in this phase contains two nearly flat bands inherent from the underlying kagome structure. The proposed complex lattice possessing a full $C_3$ symmetry can be derived from placing $\mathcal{PT}$-symmetric dimers at either the kagome sites or the honeycomb edges. 
The proposed structure probably is the simplest complex lattice that respects both the $\mathcal{PT}$ and $C_3$ symmetries.
We have also uncovered oscillatory power rotation and long-lived chiral local structures in the beam dynamics of the complex kagome lattice. The presence of nearly flat bands in a $\mathcal{PT}$-symmetric phase could lead to other novel phenomena when disorder and/or nonlinearity are taken into account, which will be the topics of future investigation.

{\em Acknowledgement.} The authors thank K. G. Makris, Y. N. Joglekar, P. G. Kevrekidis, T. Kottos, and M. Heinrich for insightful discussions. Work at Los Alamos is supported in part by the U.S. Department of Energy.

\end{document}